\documentclass[11pt,a4paper]{article}
\usepackage{jheppub}
\usepackage[T1]{fontenc}
\usepackage{amsmath,amssymb,amsfonts,amsbsy,amsthm}
\usepackage{makeidx}
\usepackage{txfonts,pxfonts}
\usepackage{bbm}
\usepackage{epsfig}
\newcommand{\be}{\begin{equation}}
\newcommand{\ee}{\end{equation}}
\newcommand{\bea}{\begin{eqnarray}}
\newcommand{\eea}{\end{eqnarray}}

\newcommand{\HH}{\mathcal{H}}
\newcommand{\E}{\mathcal{E}}
\newcommand{\TR}{\mathrm{Tr}}
\newcommand{\Fi}[1]{\mathbb{#1}} 
\newcommand{\RR}{\Fi{R}}

\usepackage{amsfonts,amsbsy,latexsym,amssymb,amscd,amstext}

\title{Tachyonic instabilities in 2+1 dimensional Yang-Mills theory and its 
connection to Number Theory}
\author[a]{Fernando Chamizo}
\author[b]{and  Antonio  Gonz\'alez-Arroyo}
\affiliation[a]{Departamento de Matem\'aticas and ICMAT \\
                 Universidad Aut\'onoma de Madrid, \\
		 Cantoblanco E-28049--Madrid, Spain}
\affiliation[b]{Instituto de F\'{\i}sica Te\'orica UAM/CSIC \\
     and  Departamento de F\'{\i}sica Te\'orica, C-15 \\
                 Universidad Aut\'onoma de Madrid, \\
		 Cantoblanco E-28049--Madrid, Spain }

		 \emailAdd{fernando.chamizo@uam.es}
	    \emailAdd{ antonio.gonzalez-arroyo@uam.es}

\abstract{We consider the $2+1$ dimensional Yang-Mills theory with gauge group $\text{SU}(N)$ on a flat 2-torus under twisted boundary conditions. We study the possibility of phase transitions (tachyonic instabilities) when $N$ and the volume vary and certain chromomagnetic flux associated to the topology of the bundle can be adjusted.  Under natural assumptions about how to match the perturbative regime and the expected confinement, we prove that the absence of tachyonic instabilities is related to some problems in number theory, namely the Diophantine approximation of irreducible fractions by other fractions of smaller denominator.
}
						    
	     
	                  \preprint{IFT-UAM/CSIC-16-088; FTUAM-16-34}

\begin{document}
\maketitle

\section{Introduction}
Yang-Mills field theory plays a fundamental role in our understanding of 
elementary particles and their interactions. This theory represents 
a challenge to physicists and mathematicians since, under a relatively
simple formulation,  it is believed to encompass a large number of phenomena
which are yet to be fully mastered  and proved. Many of these phenomena escape 
control of the standard perturbative techniques which physicists use to 
perform calculations in Quantum Field Theory. One is then faced with
the rather limited set of procedures to address this type of problems. The
most important of these follow from Lattice Gauge Theories. This
amounts to a precise definition of the corresponding path integral
formulation of these fields, and provides also a source for numerical
calculational techniques of some of  the most important
non-perturbative quantities involving these fields. Most notably we
should cite the string tension $\sigma$ which determines the
long-distance behaviour of the force induced between two sources which
couple to  these fields. In the case of electrodynamics the force falls
at large distances like the inverse distance $d$ with a coefficient given 
by the electric charge of the source $F\sim e/d^2$. In the case of non-abelian
Yang-Mills fields, the force tends to a constant $F\sim \sigma $. The
phenomenon is believed to explain the impossibility of separating the
quarks that make up a proton or a meson, and hence called quark
confinement. 

On the mathematical side, Yang-Mills  fields (non-abelian gauge fields) 
are connections on vector bundles or their associated principal fiber bundles. 
The classical and quantum dynamics of these fields is determined  by a
real valued functional on the space of connections called the Yang-Mills
functional and given by a volume integral
\be
Y(A_\mu) =\frac{1}{2 g^2}\int \TR (F\wedge *F)
\ee
where $g$ is the coupling constant and $F$ the bundle curvature 2-form.
Typically one considers the  bundle group to be a direct product of
various  SU(N) and the base space 
$\RR  \times \RR^3$ with the Minkowski metric. The most
physically interesting case being that of Quantum Chromodynamics involving an SU(3) gauge group in 3+1 dimensions, and generally accepted to be the fundamental  force underlying the strong interactions of elementary particles. However, in the Physics
literature  more general cases are considered,  either because they represent a simplification of the model or because they occur
within {\em beyond the Standard Model} scenarios, early universe
studies, etc. 

The present paper relates to problems in which  many of these
generalizations are simultaneously employed. We will be dealing with 
Yang-Mills fields with  SU(N) gauge group with large $N$, and base
manifold $\RR \times T_2$ and euclidean metric ($T_2$ stands for the
two-dimensional flat torus). Descending down to three space dimensions from the canonical 3+1 Minkowski space, produces an important simplification. However, a full understanding of this case is not available and could serve as a necessary first step in achieving the  same in four dimensions. The use of the rank of the matrices  $N$ as a variable parameter in
the game could also be an important source of information. Indeed, 
when taking $N$ to  infinity crucial simplifications occur, though 
the theory still eludes complete understanding. The limit lies also 
at  the crux of the connection of Quantum Field Theory with String
Theory. Another  ingredient in the situation that we are studying is  the formulation of the theory on a 
two dimensional compact manifold. This brings in new variables into
the game. The non-trivial topology of the base manifold is inherited 
by the bundle. This might turn out to be a bonus since the topology 
can be used as a probe of the dynamics. 

The different ingredients described in the previous paragraph 
are intimately tied to one another. The notion  of  {\em volume
independence}~\cite{EguchiKawai} introduced the idea that under certain circumstances, 
in the $N\longrightarrow \infty$ limit the physical quantities become 
independent of the torus size $l$. Thus, compactness, group rank and
bundle topology are connected. Understanding the way in which this
happens was the goal of the work published in Refs.~\cite{GPGAO}-\cite{MGPAGAMO}. The
work employed both perturbative as non-perturbative lattice techniques 
to achieve this goal. This study focused on observables which are the
eigenvalues of the Hamiltonian of the system. A conclusion was that 
the main scale of the problem is a combination of the different
parameters in the game: $x=g^2 N^2 l/(4 \pi)$. Thus when taking 
the limit of large $N$ at $\lambda=g^2 N$ fixed, the size $l$  only 
appears in this combination. The topology of the gauge fields, as
embodied in the boundary conditions for the gauge fields,  plays an 
important  role in this behaviour. `t Hooft~\cite{hooft_em} realized that, since
Yang-Mills fields transform with the adjoint representation of the
gauge group SU(N), one can consider certain topological sectors in the
space of gauge fields, which he called {\em twist sectors}. These
sectors are characterized by an integer parameter $k$ defined modulo
$N$, representing a certain discrete flux traversing the 2 dimensional
torus. The value of this integer plays an important role in the way in
which the large $N$ limit is approached. In particular, volume
independence is lost if $k=0$. The range of values of $k$ at which 
this property is preserved is still subject of investigation and was
one of the main goals of Ref.~\cite{GPGAO}. 

One can calculate the
spectrum of the Hamiltonian to a given order in perturbation theory. 
This expansion is expected to be a good aproximation at small $x$. 
If $k$ is coprime with $N$, the theory has a discrete spectrum with a
mass gap: a unique ground state separated from excited states with
energies proportional to $1/x$. The big question is whether this
separation is preserved as $x$ grows. The answer becomes doubtful  since perturbation theory predicts that the gap gets
reduced as $x$ grows. Indeed, taking perturbation theory at face value it does 
indeed predict that the gap shrinks to zero at a finite value of
$x$. If this happens we could describe this phenomenon as a {\em
tachyonic instability}, representing a certain phase transition in which  the dynamical system  behaves qualitatively distinct in different regions of $x$. 
Does this transition  occur? This is the main physics problem to which our present work is connected. The conjecture, made by one of the authors in the context of  the 4 dimensional field theory~\cite{large}, is that there is an appropriate choice
of flux $k=k(N)$ for which this transition can be avoided at each value of $N$. This result is confirmed by numerical simulations~\cite{GPGAO} at several values of $N$. 
The analysis of these numerical cases provides  an understanding of how the instability is avoided. This puts us in the right setting to formulate
the question whether the result would continue to hold at arbitrary
large values of $N$. It is at this stage  that the problem can be
translated into  a problem in Number Theory. 

In the next section we will explain the essentials of the physics  problem. For a full explanation of the context we remit the reader to Refs~\cite{GPGAO} \cite{MGPAGAMO} and references therein. Then we will reformulate the
problem in a more adequate mathematical fashion, which will allow us 
to use known results in Number Theory to address it. The paper closes 
with a summary of the conclusions and several comments about possible 
extensions and improvements.

In writing the paper we have kept in mind the possibility of a  wide spectrum of readers, according to the
interdisciplinary nature of its contents. Furthermore, 
applications of Number Theory to active physical problems are scarce. 
The work itself serves to exemplify the potential fruitfulness 
of these cross collaborations.

\section{Statement of the problem}
In this section we will formulate in a more precise fashion what the 
physical problem that we are addressing is about. As mentioned earlier 
we are considering  SU(N) Yang-Mills fields 
living  in a 2-dimensional torus times the real line
$\mathbb{R}\times T_2$. For simplicity we take this torus to be a
square torus with euclidean metric and having the same length $l$
in both directions. The non-periodic direction can be considered the
euclidean time direction. At the classical level gauge fields are
connections in an SU(N) vector bundle. The bundle is given in terms 
of patches and transition matrices. As is typical in the Physics
literature we will work with a single coordinate patch
covering the full torus and a trivialization of the bundle. The 
gauge fields are then specified by the connection 1-form in this 
patch: the vector potential $A_\mu(x_0,x_1,x_2)$, which takes values in the Lie 
algebra of the group. 
As usual, the time coordinate is labeled with $0$ and placed in the first place.
The non-trivial character of the bundle arises 
through the boundary conditions imposed on the vector potential:
\bea
A_\mu(x_0,x_1+l, x_2)= \Gamma_1 A_\mu(x_0,x_1,x_2) \Gamma_1^\dagger \\
A_\mu(x_0,x_1,x_2+l)= \Gamma_2 A_\mu(x_0,x_1,x_2) \Gamma_2^\dagger 
\eea
where $\Gamma_i$ are SU(N) matrices satisfying 
\begin{equation}\label{t_eater}
\Gamma_1 \Gamma_2 = e^{2 \pi i k/N} \Gamma_2 \Gamma_1
\end{equation}
with $k$ an integer defined modulo $N$. 
These relations derive from the most general twisted boundary conditions satisfied by the transition matrices $\Omega_1$, $\Omega_2$
(cf. \cite{hooft_em} \cite{hooft_q})
\begin{equation}\label{twisted}
\Omega_j(x+L\vec{e}_l)\Omega_l(x)
=
e^{2\pi i n_{jl}/N}
\Omega_l(x+L\vec{e}_j)\Omega_j(x)
\qquad\text{where}\quad
n_{jl}=-n_{lj}.
\end{equation}
We will restrict ourselves
to the case in which $k=n_{12}$  is coprime with $N$. The different values of 
$k$ label topologically inequivalent sectors, called {\em twist
sectors} by `t Hooft. 
He also explained the physical 
interpretation of this quantity as a certain discrete chromomagnetic
flux. The reader is addressed to the appropriate references~\cite{hooft_em} \cite{hooft_q} \cite{vanbaal2} \cite{peniscola} for a 
more detailed explanation of the physical and mathematical
interpretation of the twist. The solutions $\Omega_1(x)=\Gamma_1$, $\Omega_2(x)=\Gamma_2$ 
of Eq.~\eqref{twisted} with $\Gamma_1$ and $\Gamma_2$ constant matrices as in Eq.~\eqref{t_eater} 
are called twist eaters (see \cite{vanbaal} and \cite{lebedev} for their general form).

Our interest is focused in the quantum version of this field theory. 
According to the postulates of quantum physics the set of
(gauge-invariant) states of the system are given by rays in a    
Hilbert space $\HH$. Our main interest is the study of the spectrum 
of the Hamiltonian operator $H$, which  acts on this space. In the
alternative path-integral quantization the spectrum of the Hamiltonian 
determines the exponential fall off of correlation functions of gauge
invariant observables at different euclidean times 
\be
\langle O(0) O'(x_0) \rangle \longrightarrow \langle O(0) \rangle
\langle O'(0) \rangle + A e^{-x_0 E} + \ldots
\ee
where $E$ is an  eigenvalue of the Hamiltonian. It is worth mentioning here that 
it is enough to consider as our gauge invariant observables the algebra 
generated by spatial Wilson loops: the trace of the parallel transport 
matrices along a closed path $\gamma$ on the 2-torus. 

As first understood  by `t Hooft~\cite{hooft_em}, the formulation of the
theory on the torus introduces a new  ingredient in the game: {\em
center symmetry}. This is a set of transformations,  forming a group 
isomorphic to $Z_N^2$, which commute with the Hamiltonian. Under these 
transformations the Wilson loops associated to paths with non-zero winding
transform multiplicatively.  The irreducible representations of this
group are labelled by a two dimensional vector of integers defined
modulo $N$: $\vec{e}=(e_1,e_2)$. Hence, the Hilbert space can be
decomposed into a direct sum of spaces associated to each irreducible 
representation:
$$ \HH = \bigoplus_{\vec{e}} \HH_{\vec{e}} $$
The integer-vector  $\vec{e}$ is called the (chromo-)electric flux 2-vector.
It is clear that the Hamiltonian operator does not change the value of 
the electric flux. We may then write 
$$ H_{\vec{e}}: \HH_{\vec{e}} \longrightarrow \HH_{\vec{e}} $$
and our interest is centered  on the  study of the spectrum of the Hamiltonian $H_{\vec{e}}$ in each of these sectors.

After this introduction we are now in position of explaining what is
the main goal of our physics program. We would like to compute and 
interpret the lowest energy levels of the Hamiltonians $H_{\vec{e}}$. 
These are given by the following real quantities:
$E_p(\vec{e},\lambda,l,N,k)$, where $p$ is a non-negative integer index listing the
eigenvalues in each sector in increasing order. A  very important 
part of our program, connected with the interpretation goal, is that of 
determining the dependence of the energies $E_p$ on the two real parameters $\lambda$ and $l$
and the two discrete parameters $k$ and $N$.

By definition the ground state in the
$\vec{e}=0$ sector is called the vacuum and its energy is set to 0.
This is part of the renormalization program, necessary to make sense 
of our computational program.  The lowest eigenvalue of $H_{\vec{e}}$ 
in the remaining sectors is called the energy of the electric flux 
sectors. Its value depends on all the parameters that describe the system
$E_0(\vec{e},\lambda,l,N,k)$, where $\lambda$ is `t Hooft coupling
constant: $\lambda=g^2 N$. It is important to realize that in 2+1
dimensions $\lambda$ has dimensions of energy. Thus, if we apply
dimensional analysis one must have
$$ E_0(\vec{e},\lambda,l,N,k)= \lambda \E(\vec{e}, \lambda l, N, k) $$
Our goal is to determine the function $\E$. In addition, we are also
interested in the first excited energy (the second lowest eigenvalue) of 
$H_{\vec{e}=0}$: $ E_1(\vec{0},\lambda,l,N,k)$. This is called the lowest
glueball mass. 

There is only one way  in which physicists know how to compute the
energy levels without making uncontrolled assumptions: lattice gauge
theory~\cite{wilson}. This, so-called {\em first principles}
calculation,
is based on formulating the problem as that of computing expectation
values in a probability measure over the space of  a finite number 
of $\text{SU}(N)$ matrices. The philosophy is to discretize the  
$\mathbb{R} \times T_2$ space into a cubic lattice 
in which the points are separated by a distance $a$: the 
lattice spacing. The number of points in the spatial directions 
is given by $L=l/a$. In principle there are infinite points 
after discretizing the non-compact directions, however the 
energy levels can be determined with arbitrary precision 
even if we consider this direction finite as well, provided 
its length is large enough. The quantum field theory results, 
including our desired energies, are obtained after taking the 
limit $a\longrightarrow 0$. This limit has to be accompanied 
by an appropriate tuning of the probability density, a process 
known as {\em renormalization}. At the end of the day physical 
results are obtained which do not depend on the details of the 
discretization procedure. This statement has not been proved 
rigorously, so that from the mathematical standpoint can be 
considered a conjecture. However, there is a large number of 
results of all types that gives credibility to the conjecture. 
Indeed, proving this conjecture can be considered a possible 
ingredient in solving one of the millenium problems stated 
by the Clay foundation. 

In previous papers, one of the present authors and collaborators 
used lattice gauge theory to compute the $E_0(\vec{e},\lambda,L,N,k)$
quantities for certain values of the arguments~\cite{GPGAO} \cite{MGPAGAMO}. The study 
of the glueball masses $E_1(\vec{0},\lambda,l,N,k)$ is currently 
underway~\cite{prepar}. Remember, however, that our goal is also 
to interpret and understand the results, and for that purpose 
the dependence on the different parameters is crucial. 

As stated earlier the dependence on the size of the torus comes 
in the combination $l\lambda$, which is a dimensionless parameter. 
A well-known calculational technique in quantum field theory 
is perturbation theory, which provides an (asymptotic) expansion of
the physical  quantities in powers of $\lambda$. Obviously the
expansion becomes a better approximation for small $l\lambda$. The
leading term is very simple to obtain, since at that order the system 
behaves like a quantum system of free massless bosons: the gluons. Each gluon 
has an energy given by $|\vec{p}|$, the modulus of its momentum
$\vec{p}=(p_1,p_2)$. Because of the boundary conditions the momentum
of the gluons  is quantized as follows:
\be
p_j=\frac{2 \pi  n_j}{N l} 
\ee
where the $n_j$ are integers. There is a connection between these
integers and the electric flux quantum number as follows:
\be
\vec{n}= k \vec{e}_\bot + N \vec{m}
\ee
where $k$ is the magnetic flux,  $\vec{e}_\bot=(-e_2,e_1)$ and $\vec{m}$ is a 2-vector of integers. 
Hence, the minimal energy  is given by $2 \pi/(Nl)$ and corresponds 
to the electric flux sectors $(\bar{k},0)$ and $(0,\bar{k})$, where 
$\bar{k}k= \pm 1 \bmod N$ with $|\bar{k}|\le N/2$.
If we introduce the common notation in number theory $\|x\|$ to denote the distance of the
real number $x$ to the nearest integer, then $N\|k/N\|$ is the representative
of the class of $k$ or $-k$ modulo $N$ that fits the interval $[0,N/2]$. With this notation,
for a generic value of the electric flux $\vec{e}$, the minimal energy is given by
\begin{equation}
\frac{2\pi}{Nl}
\sqrt{N^2\|ke_1/N\|^2+N^2\|ke_2/N\|^2 }
=
\frac{2\pi}{l}
\sqrt{\|ke_1/N\|^2+\|ke_2/N\|^2 }\equiv \frac{2\pi}{l} \|k\vec{e}/N\|_2  .
\end{equation}
where in the last equality we have introduced a new rather natural notation.

For large values of $l\lambda$  we enter a non-perturbative 
domain where we lack calculational techniques other than the lattice. 
Nevertheless, this is also the regime of non-compact space-times 
($l \longrightarrow \infty$)  in which most physical results
concentrate. The concept of quark confinement leads to 
the expectation that the energies of the non-zero fluxes grow linearly 
with the torus length $l$. Combining this with  dimensional analysis, it
leads to 
\be
\E(\vec{e},\lambda l,N,k) \longrightarrow \bar{\sigma}(\vec{e},N)
\lambda l
\ee
where we used the idea that for large sizes the magnetic flux 
value $k$ (entering only in the boundary conditions) becomes irrelevant. The quantity $\bar{\sigma}(\vec{e},N)$
is the well-known string tension measured in $\lambda^2$ units. The
name reflects the interpretation of the phenomenon, as the formation 
of a chromo-electric flux tube (a thick string) carrying a certain 
energy per unit length. This phenomenon being the electric-magnetic 
dual of the Meissner effect in superconductors.

Now the problem is focused upon understanding how the energies evolve 
from the $1/l$ behaviour of perturbation theory to the linear  $l$ dependence
typical of large volumes. The $N$ dependence could serve to clarify
this transition. One crucial question would be whether in the large 
$N$ limit the energies become discontinuous or non-analytic in $l\lambda$. 
These type of situations is labelled as a large $N$ phase transition 
in the literature. The purpose of the present  work is precisely that
of advancing in the resolution of this question. 

When stuying the large $N$ behaviour of the system a crucial ingredient 
is that of investigating the phenomenon of {\em volume
independence}~\cite{EguchiKawai}. This means that in the limit of
large $N$ the physical results become independent on the torus size $l$. 
Whether this is true and how this situation is approached is still a
subject of debate. From early times it  is known that using twisted boundary conditions is crucial for  the implementation of this idea~\cite{twisted2} \cite{twisted1}. In the 4 dimensional euclidean setting it has been 
argued recently~\cite{large} that the phenomenon indeed takes place provided the magnetic
flux $k$ and its congruency inverse $\bar{k}$ are scaled linearly 
with $\sqrt{N}$, as $N$ grows. Our previous work on the ground state energies 
of the non-vanishing electric flux sectors of the 2+1 dimensional
system provided numerical evidence that a similar situation (with $N$ replacing $\sqrt{N}$) also occurs here~\cite{GPGAO}~\cite{MGPAGAMO}. 

What are the reasons suggesting the possibility of a large $N$ phase transition? These come mainly from the next-to-leading 
order calculation of the energies in  perturbation theory.  The contribution turns out to be negative and physically can be interpreted as a self-energy for the gluon. Following the notation of Ref.~\cite{GPGAO} we will write the perturbative contribution as follows:
\be
\label{energysq}
\E^2(\vec{e},\lambda l,N,k)=\frac{\phi_0(\vec{e},N,k)}{4 x^2}- \frac{G(\vec{e},N)}{x} + \ldots
\ee
where 
\be
x=\frac{\lambda l N}{4 \pi}
\ee
\be
\phi_0= N^2 ( || ke_1/N||^2+ || ke_2/N||^2)= N^2 ||k \vec{e}/N||_2^2
\ee
and the self-energy is given by 
\be
G(\vec{e},N)=-\frac{1}{16\pi^2}\int_0^\infty \frac{dt}{\sqrt{t}} \large( \theta_3^2(0,it)
-\theta_3(e_1/N,it)\theta_3(e_2/N,it)-\frac{1}{t}\large) 
 \ee
 where $\theta_3(z,\tau)$ is Jacobi theta function, defined 
 by the  series
\be
\theta_3(z,\tau)
=
\sum_{n=-\infty}^\infty
\text{exp}\big(
\pi in^2\tau+2\pi inz
\big)
\ee
For the numerical evaluation of the self-energy one can use this series  when $t$ is large. For small values it is advantageous to use the  modular relation
\be
\theta_3\big(\frac{z}{\tau},-\frac{1}{\tau}\big)
=
(-i\tau)^{1/2}e^{\pi iz^2/\tau}
\theta_3(z,\tau)
\ee
that follows easily from Poisson summation formula \cite[(52)]{siegelb}.

 We recall that the first term on the right-hand side in Eq.~\eqref{energysq} is the momentum square of the gluon in $\lambda$ units. Thus, by analogy with the relativistic dispersion relation, the remaining terms can be thought of as giving the mass square. If $G$ is positive this would mean a negative mass square, a situation often described as {\em tachyonic}. Furthermore, if we neglect all the higher corrections appearing as dots in the right-hand side of Eq.~\eqref{energysq}, then for $x\ge\bar{x}$ given 
 by 
 \be
 \bar{x}=\frac{\phi_0(\vec{e},N,k)}{4 G(\vec{e},N)}
 \ee
 the energy square becomes negative. This situation makes no sense and it signals the breakdown of perturbation theory and the system entering a different phase in which the vacuum has a non-vanishing condensate. This is precisely the situation that is described as {\em tachyonic instability}. There is numerical evidence that this instability does indeed take place for some cases of $\vec{e}/N$ and $k$.  Our main goal can  now be stated clearly: we want to see if  for every $N$ there  are choices of  $k$ for which the instability can be avoided for all $\vec{e}$. Previous numerical work obtained  at certain values of $(k,N)$ suggests that this is the case.  However, we would like to know if the situation survives the large $N$ limit.

 To analyze this problem further, let us examine the previous arguments suggesting the instability. The weak point is the fact that we neglected corrections which have higher positive powers of $x$ and dominate for large $x$. There is one case, however, in which the  argument in favour of instabilities becomes compelling and that is when $\bar{x}$ becomes very small as $N$ goes to infinity. This is actually happening in several cases because $G(\vec{e},N)$ has indeed a pole singularity 
 \be
G(\vec{e},N)\sim \frac{1}{16 \pi^2 ||\vec{e}/N||_2}+ R(\vec{e}/N)
\ee
where $R$ is a positive definite regular function which vanishes at the origin and numerically rather small. Indeed, it is bounded as follows
\be
0.01< \frac{R(\vec{e}/N)}{||\vec{e}/N||_2^2} <0.02
\ee 
Hence,  for the sake of studying the possible instability it is a good approximation to consider only the pole part in estimating $\bar{x}$:
\be
\bar{x}\sim 4 \pi^2 N^2\ ||k \vec{e}/N||_2^2\ ||\vec{e}/N||_2
 \ee
 Now given the definition it is easy to see that the right-hand side  is bounded from below by 
 \be
4 \pi^2 N^2\ (||k e_1/N||^2\ ||e_1/N||+||k e_2/N||^2\ ||e_2/N||) 
\ee
Hence, if we want to find a lower bound of $\bar{x}$ for all values of $\vec{e}$ it is enough to find the minimum for electric fluxes of the form $(e,0)$. Our goal then is that of tuning $k$ in order to maximize this minimum. The result will be called $a(N,2)$
\be \label{a_case2}
a(N,2)=
\max_k\min_e
N^2\|ke/N\|^2\|e/N\|
\ee
This,  sets our first mathematical goal, that of studying whether $a(N,2)$ can be bounded from below uniformly as $N$ runs over all integers beyond a certain one.  

Obtaining such a bound eliminates the argument in favour of a tachyonic instability in the perturbative region. However, it could still happen that at finite values of $x$ one of the minimum flux energies crosses zero. To eliminate this possibility we need to have some control about the possible additional contributions for higher values of $x$. We have already  
commented about the behaviour at large $x$ arising from the phenomenon of quark confinement. Indeed, it is also possible to 
describe the expected subleading terms that govern the approach towards the confinement regime. These follow from the effective string picture description of the flux tube formation. The leading correction  has a universal character and is often referred as Luscher term~\cite{luscher} \cite{Luscher:1980ac}. Thus, using the same notation as for the perturbative part we can express the expected behaviour of the flux energies for large values of $x$
\be
\label{largex}
\E^2(\vec{e}, \lambda L, N, k) = \frac{1}{4}\tau^2\tilde{\sigma}^2(\vec{e}/N)
x^2  
-\frac{\tau }{24}\chi(\vec{e},N,k)
+\ldots
\ee
where we have parameterized the string tension  as 
follows
\be
\bar{\sigma}(\vec{e},N)=\frac{\tau N}{8 \pi}\tilde{\sigma}(\vec{e}/N)
\ee
where $\tau$ is a numerical coefficient whose value has been determined numerically to be close to 1, as predicted by Nair~\cite{KaKiNa} (see also \cite{nair} \cite{BrTe}). 
The function $\tilde{\sigma}(\vec{e}/N)$
describes what is known as the $k$-string spectrum and for small value of the argument  goes like $\|\vec{e}/N\|$.
The correction term is $x$-independent. The dependence on $\vec{e}$ is controlled by the function  $\chi(\vec{e},N,k)$ which goes to 1 as $||\vec{e}/N||_2$ goes to zero. The normalization is fixed by the numerical value of the Luscher term for the string. 

\begin{figure}[!htb]
\centering
\includegraphics[scale=1.]{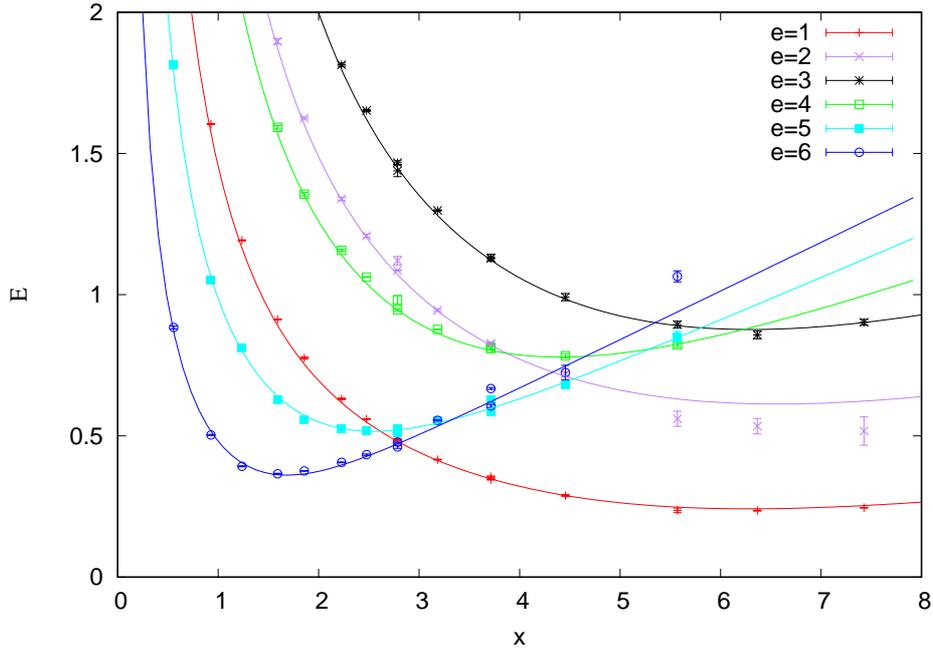}
\caption{Comparison of flux energies $\mathcal{E}$ with Eq.~\eqref{totalfunc} as a function of $x$ for the case $N=17$, $k=3$}
\label{fig:ener}
\end{figure}

Some comments about Eq.~\eqref{largex} are interesting. The first is that higher order string-like corrections vanish for the particular case of the Nambu-Goto string. The second  comment is that the constant (Luscher) term has the same $x$ dependence as a possible $\lambda^2$ contribution in perturbation theory. Finally, we point out that, given that $\tilde{\sigma}(\vec{e}/N)$ is an increasing function of its argument, the right-hand side of Eq.~\eqref{largex} acquires a certain magnitude at a lower value of $x$ the bigger the value of the corresponding electric flux. The numerical results of Ref.~\cite{GPGAO} also show an earlier (in terms of $x$) departure from the perturbative behaviour for higher electric fluxes.  This suggested the authors to try to compare the numerical data on the energies with the formula 
\be
\label{totalfunc}
\E^2(\vec{e}, \lambda L, N, k) = \frac{\phi_0(\vec{e},N,k)}{4 x^2}- \frac{G(\vec{e},N)}{x} -\frac{\tau }{24}\chi(\vec{e},N,k)+\frac{1}{4}\tau^2\tilde{\sigma}^2(\vec{e}/N) x^2
\ee 
obtained by adding the perturbative terms, dominant for small $x$, with the confining terms, valid for large ones. Curiously the formula describes rather well the behaviour of the data even at intermediate values of $x$. This is exemplified in Fig.~\ref{fig:ener},  in which the energies for various electric fluxes $\vec{e}=(e,0)$ for $N=17$ and $k=3$ are plotted as a function of $x$. The data points are from Ref.~\cite{MGPAGAMO} \cite{prepar}. The continuous lines having the same colour as the points result from using Eq.~\eqref{totalfunc} with $G(\vec{e},N)$ given by the pole, $\frac{\tau }{24}\chi(\vec{e},N,k)=0.03$,  and    $\tau\tilde{\sigma}(\vec{e}/N)=1.2\sin(\pi e/N)/\pi$.  
It is clear that the energies for each of the electric fluxes follow the pattern given by the equation. The qualitative agreement is not spoilt by small variations of $\tau$ or the constant term. The same pattern is present for other values of $k$ and $N$. However, all the tested values of $N$ are prime numbers. This is done for simplicity as well as to avoid the existence of non-trivial subgroups of $Z_N$.  It is unclear if the formula also describes the data when $e$ and $N$ have common divisors.  Hence, at least for prime $N$, we have  a rationale to understand the behaviour of the energies which allows us to predict  whether tachyonic instabilities would develop at intermediate values of $x$.

Now the question becomes: assuming the validity of the formula, can we prove that for every $N$ there is a value of $k$ such that all energies 
$\E^2(\vec{e}, \lambda L, N, k)$ are larger than zero?
To work out the analytical conditions following from this assumption 
we first realize that the condition can be translated into a condition on the quantity $Z=\tilde{\sigma}(\vec{e}/N)\sqrt{\phi_0(\vec{e},N,k)}$. As a matter of fact we can approximate 
$\tilde{\sigma}(x)=||x||_2$, so that we can write
\be
Z= N ||\vec{e}/N||_2 ||k\vec{e}/N||_2 
\ee
Defining $y=\sqrt{\phi_0(\vec{e},N,k)}/x$, the energy square can be written 
\be
\E^2(\vec{e}, \lambda L, N, k)=\frac{y^2}{4}-\frac{y}{16 \pi^2 Z}-\frac{\tau}{24}\chi(\vec{e},N,k)+\frac{\tau^2 Z^2}{4 y^2}
\ee
where we have used the pole form of the self-energy. 
Now we can minimize this function with respect to $y$ and compute the value of the function at the minimum. This can be done analytically since the equation for the minimum is a quartic equation, whose single real and positive root can be substituted back into the function of the energy. The result for the minimum energy 
takes the form 
\be
\bar{\cal E}^2=f(Z,\tau)-\frac{\tau}{24}\chi(\vec{e},N,k)
\ee
All that we need to know about the function $f$ is that it is a monotonously growing function of $Z$ in the region of interest and with a slight dependence on $\tau\sim 1$. Indeed, $\chi(\vec{e},N,k)$ is also approximately equal to 1, although it does not play any role in our reasoning. The minimum energy would then cross zero at an specific value of $Z$, which we will call $Z_0$. Varying the parameters within reasonable limits we obtain values of $0.09\le Z_0\le 0.12$. The conclusion  then is that we would be able to avoid tachyonic instabilities at a given value of $N$ and $k$ provided  $Z>Z_0$ for all values of the electric flux. Again this minimum condition is saturated by electric fluxes of the type $(e,0)$. 
Hence, our original problem then reduces to proving that for any $N$, one has
\be \label{a_case1}
Z_0< a(N,1)=\max_k\min_e
N\;\|ke/N\|\;\|e/N\|
\ee
It is this condition that will be explored in the next section using known results in number theory.

\section{Connection to Number Theory results}

According to our previous considerations, the existence of tachyonic instabilities is related
to the behavior of the sequence
$\big\{a(N,n)\big\}_{N=1}^\infty$ defined as
\begin{equation}\label{a_Nn}
a(N,n)=
\max_k\min_e
N^n\|ke/N\|^n\|e/N\|
\end{equation}
where $k$  runs over integers relatively prime to $N$ and $e\ne 0 \bmod N$. 
As we have seen in Eq.~\eqref{a_case2} and Eq.~\eqref{a_case1}, the value $n=2$ appears when studying   tachyonic instabilities in the perturbative regime and $n=1$ corresponds to the general case.

Some of the most natural problems regarding this sequence and their physical motivations are in the following list:
\begin{enumerate}\setlength{\itemsep}{1pt}
 \item 
 If a universal lower bound holds for $n=2$, then the existence of instabilities at the level of perturbation theory is disproved.
 \item 
 If $a(N,n)$ has a large enough universal lower  bound for $n=1$ as $N\to\infty$, then one can show that it is possible to avoid the existence of instabilities in the model. 
  \item
 An algorithm to find a $k$  in Eq.~\eqref{a_Nn} perhaps not reaching the maximum  but establishing a good lower bound for $a(N,n)$ would  give a way of finding for each $N$ a twist $k$ to avoid instabilities. 
 \item
 The possibility of bounds of this kind for a sequence $N_j\to\infty$ (and the corresponding $k_j$) would establish a form of defining a large $N$ limit of the model.
 \item
 An algorithm to restrict the possibilities for $k$ and $e$ in Eq.~\eqref{a_Nn} would be very convenient to carry out numerical studies of the stability of the model.
\end{enumerate}

In what follows we will address the points listed above. For that purpose we will introduce some terminology which will allow us to reformulate our problem in a number theoretical fashion. 

First we recall that  the Farey sequence $\mathcal{F}_N$ (rather a set in our case)  is the set of irreducible fractions in $[0,1]$ with denominator at most~$N$. Then we can rewrite 
\begin{equation}\label{a_NF}
a(N,n)=
N^{n-1}
\max_\alpha\min_{p/q\le 1/2}
q^{n+1}
\Big|
\alpha-\frac pq\Big|^n
\qquad\text{where}\quad
\alpha\in \mathcal{F}_N-\mathcal{F}_{N-1}
\quad\text{and}\quad 
\frac pq \in \mathcal{F}_{N-1}.
\end{equation}
In other words, $\alpha$ is an irreducible fraction with denominator exactly $N$ while $p/q$ has a denominator $q$  strictly smaller.

To deduce Eq.~\eqref{a_NF} from Eq.~\eqref{a_Nn}  we should identify 
$\|e/N\|=q/N$ and $\alpha=\|k/N\|=l/N$.  Then one has \begin{equation}
a(N,n)= \max_l\min_q
N^{n-1} q\, \Big\|\frac{lq}{N}\Big\|^n 
\end{equation}
which shows that $a(N,n)$ is invariant under 
\begin{equation}
k\mapsto N-k
\qquad\text{and}\qquad
e\mapsto N-e \quad .
\end{equation} 
Hence, we can restrict ourselves to  $0<k,e< N/2$, and then the integers $l$ and $q$ would coincide with $k$ and $e$ respectively. 
In addition one can rewrite 
\begin{equation}
\Big\|\frac{lq}{N}\Big\|= \min_p \Big|\frac{lq}{N}-p\Big|=q \min_p \Big|\alpha-\frac{p}{q}\Big| 
\end{equation}
which rather easily leads to Eq.~\eqref{a_NF}

The previous formulation drives us to the question of approximating a real number by an irreducible fraction (with fixed denominator), and this calls for the use of continued fractions. This constitutes a  classical topic in number theory. See for instance the reference \cite{miller} for the properties stated below. 

Any rational number can be expressed as a continued fraction
\begin{equation}
 [a_0; a_1,a_2,\dots, a_M]
 :=
 a_0+1/\Big(a_1+1/\big(a_2+1/(a_3+\dots)\big)\Big)
 \quad
 \text{where $a_0\in\mathbb{Z}$ and $a_1,a_2,\dots, a_M \in\mathbb{Z}^+$}.
\end{equation}
In fact this representation is unique for nonintegral numbers imposing $a_M>1$. For instance, $5/11=
[0, 2, 5]$
and $77/103=
[0, 1, 2, 1, 25]$. 

The fractions 
\begin{equation}
\frac{p_n}{q_n}=[a_0; a_1,a_2,a_3,\dots, a_n]
\end{equation}
are called convergents of the continued fractions (assumed irreducible).  Notice that if $M$ is the number of terms in the continued fraction then  $p_M=l$ and $q_M=N$.
Very often it is also defined $p_{-1}=1$ and $q_{-1}=0$. In this way, we have the recurrence formulas
\begin{equation}\label{mform}
p_j=a_j p_{j-1}+p_{j-2}
\qquad\text{and}\qquad
q_j=a_j q_{j-1}+q_{j-2}
\qquad\text{for $j\in\mathbb{Z}^+$.}
\end{equation}
These are the same formulas that those for the Euclidean algorithm. One can prove that the so-called partial quotients $a_j$ are actually the successive quotients obtained when one applies the Euclidean algorithm  to the numerator and denominator of the initial rational number in its irreducible form.

The key property that we are going to use is that the convergents give optimal approximations. It means that if $p_j/q_j$ are the convergents of $\alpha$, then 
\begin{equation}\label{ska}
|q_j\alpha-p_j|
 =
 \min_{p/q\in\mathcal{F}_{Q}}
 |q\alpha-{p}|
\qquad
\text{for any }\
q_j\le Q<q_{j+1}.
\end{equation}
This is valid, even if $\alpha\in\mathbb{R}$ is not rational, in this case we have an infinite continued fraction. Since  
$q\ge q_j$ this also holds for $q^{n+1}|\alpha-p/q|^n$ in the same interval. 

Thus property Eq.~\eqref{ska} proves that we can restrict ourselves to consider $p/q$ as a convergent of $\alpha$ in Eq.~\eqref{a_NF}. This is a major advance in the fifth point of the list above with respect to a brute force search for the minimum. The number of steps of the Euclidean algorithm is $O(\log m)$ where $m$ is the minimum of the initial numbers. Then the computation of $a(N,n)$ examining all the convergents requires at most $ O(N\log N)$ steps. 
Recall that $k= N\alpha$ and $e=q$, then once the magnetic flux $k$ is fixed, there are very few electric fluxes to be checked. 

On the other hand, if we expect a bound $a(N,n)>c_0$ then it is not necessary to consider $\alpha$ such that
$|\alpha-p/q|\le
c_0^{1/n}q^{-1-1/n}N^{-1+1/n}$.
In other words, we can omit the values of $k$ in the intervals 
$|k-Np/q|\le (c_0 N)^{1/n}q^{-1-1/n}$. The interval is larger for small values of $q$. A method in the direction of the fifth point of the list is to perform a preliminary sieve of the values of $k$ choosing $p/q\in\mathcal{F}_Q$ with $Q$ not very large. 
Indeed, we have used the methods described above to make a scan of the minimum values of $a(N,n)$ for the first few thousands prime values of $N$. We will comment on the results later.
%
%

It is possible to give an alternative formula  \cite[\S7.5]{miller} for the left hand of Eq.~\eqref{ska}:
\begin{equation}\label{eska}
 q_j\alpha-p_j=\frac{(-1)^j}{\alpha_{j+1}'q_j+q_{j-1}}
\qquad\text{with}\quad
\alpha'_j=[a_j;a_{j+1},\dots, a_M ],
\end{equation}
where, as before,  $[a_0;a_1,a_2,\dots, a_M]$ is the continued fraction of $\alpha$. 
By Eq.~\eqref{a_NF} and Eq.~\eqref{ska}
\begin{equation}\label{a_err}
a(N,n)=
N^{n-1}
\max_{\alpha}
\min_{j<M}
\frac{q_j}{\big(\alpha_{j+1}'q_j+q_{j-1}\big)^n}
\qquad\text{with }\ 
\alpha'_j=[a_j;a_{j+1},\dots ,a_M].
\end{equation}
After this long introduction we are now ready to address the first two points of our previous list. Let us begin with the $n=2$ case.

\

\paragraph{\large The case n=2 case.}
For prime values of $N>2$ (the $N=2$ case is trivial) we are going to prove that
\begin{equation}\label{case2}
a(N,2)
>\frac{3}{\pi^2}\big(1-N^{-1}\big).
\end{equation}
The method below gives a
slightly  better bound. 

Restricting $k$ to the interval $(0,N/2)$,  we can define a function  $F(k)$ and re-express Eq.~\eqref{a_NF} in terms of it as follows:
\begin{equation}\label{f_orig}
F(k)
=
N \min_{p/q}
q^3
\Big| \frac kN-\frac pq\Big|^2
\qquad\text{so that}\quad
a(N,2)=
\max_{0<k<N/2} F(k). 
\end{equation}
We can also express $F(k)$ as follows 
\begin{equation}\label{f_altern}
F(k)=\min_{0<m<N/2} m^2 \frac{q(k,m)}{N}\equiv \min_{0<m<N/2} m^2  \Big\|\frac{\bar{k}m}{N}\Big\|
\end{equation}
where we recall that $\bar{k} k= 1 \bmod N$. 

Given $k$, say that the minimum appearing in \eqref{f_altern} is attained for a particular value of $m$. We define the sets
\begin{equation}
\mathcal{C}_m
=
\big\{ 0<k<N/2\; :\; q(k,m)=\frac{N F(k)}{m^2} 
\big\}
\end{equation}
Clearly, all the values of $k$ must belong to at least one of the sets so that 
\begin{equation}\label{union}
\sum_m
\#\mathcal{C}_m
\ge
\#
\Big(
\bigcup_m
\mathcal{C}_m
\Big)
\ge
\#\{0<k<N/2\}=(N-1)/2,
\end{equation}
where $\#A$ stands for the cardinality of the set $A$.
On the other hand, if $k\in \mathcal{C}_m$ then
\begin{equation}\label{ant317}
q(k,m)=\frac{N F(k)}{m^2} \le \frac{N a(N,2)}{m^2}.
\end{equation}
Let us now define the sets 
\begin{equation}
\mathcal{D}_m
= 
\big\{ 0<q<N/2\; :\; \exists k\in \mathcal{C}_m\  / \ q(k,m)=q
\big\}
\end{equation}
From Eq.~\eqref{ant317} we conclude that $\#\mathcal{D}_m\le 
{N}a(N,2)/{m^2}$.

The last step is to realize $\mathcal{D}_m$ has the same cardinality as $\mathcal{C}_m$. To see this we  notice that $\bar{k}=\bar{m}q\pmod{N}$, where $\bar{m}$ is the congruent inverse of $m$.
 From these observations and~Eq.~\eqref{union}, we get
\begin{equation}
(N-1)/2\le
\sum_{0\ne m <N/2}
\frac{N}{m^2}
a(N,2)
<
a(N,2) N
\sum_{m=1}^\infty
\frac{1}{m^2} = a(N,2) N \zeta(2)
\end{equation}
Substituting the value of $\zeta(2)$ gives Eq.~\eqref{case2}. 
Our proof succeeds in avoiding the existence of tachyonic instabilities in the perturbative region for prime $N$.

A similar analysis in the case $n>2$,  produces
\begin{equation}
\frac{N-1}{2}\le
\sum_{0\ne m<N/2}
\frac{N}{m^n}
a(N,n)
\le
\zeta(n)
N
a(N,n).
\end{equation}
On the other hand, 
the last convergent of $\alpha$ different from itself and $\alpha$ are consecutive Farey fractions. If we take it as $l/q$ in Eq.~\eqref{a_NF}, we have $|\alpha-l/q|=(qN)^{-1}$ and $a(N,n)<1/2$ because $q<N/2$. In this way, we have the upper and lower bounds 
\begin{equation}\label{}
\frac{1}{2\zeta(n)}\big(1- N^{-1}\big)
<a(N,n)
<\frac 12
\qquad\text{for $n>1$}.
\end{equation}
If $N$ and $n$ go to infinity, $a(N,n)$ tends to be $1/2$. For $n=2$, we have  $2\zeta(n)=\pi^2/3$ 
that gives~Eq.~\eqref{case2}.

\paragraph{\large The  n=1 case.}
Firstly we are going to show that in the case $n=1$ the optimal values of $N$ to avoid tachyonic instabilities are the Fibonacci numbers. 

Recall the (extended) Fibonacci sequence
$\{F_k\}_{k=0}^\infty=(0,1,1,2,3,5,8,\dots)$
defined by the recurrence $F_{j+2}=F_{j+1}+F_{j}$.
Assume that $N\ge 5$ is a Fibonacci number, say $N=F_J$. Thanks to Eq.~\eqref{mform}, it is easy to see that
\begin{equation}
[0;2,\overbrace{1, {\dots\dots\dots},1}^\text{$J-5$ times},2]
=
\frac{F_{J-2}}{F_{J}}
=
\frac{F_{J-2}}{N}.
\end{equation}
Let us call this number $\alpha_0$. We are going to check that
$a(N,1)=\alpha_0$. The convergents are
$0/1,1/2,1/3,2/5,\dots, F_{J-4}/F_{J-2}$ and $\alpha_0$.
In the same way, $\alpha_1'=1/\alpha_0$, $\alpha_j'=F_{J-j}/F_{J-j-1}$.
Using the properties of the Fibonacci sequence, namely \cite[p.89, 47($r=1$)]{koshy}
and \cite[p.88, 18($n=2$)]{koshy}, we have for $0<j\le J-4$
\begin{equation}
q_j-\alpha_0\big(\alpha_{j+1}'q_j+q_{j-1}\big)
=
q_j-\alpha_0\Big(\frac{F_{J-j-1}}{F_{J-j-2}}F_{j+2}+F_{j+1}\Big)
=
F_{j+2}-\frac{F_{J-2}}{N}\frac{N}{F_{J-j-2}}
=
\frac{F_{J-j-4}F_j}{F_{J-j-2}}\ge 0
\end{equation}
which is still valid for $j=0$ with equality. 
Hence
for $\alpha=\alpha_0$ the minimum in Eq.~\eqref{a_err} is reached for $j=0$ giving $a(N,1)\ge \alpha_0$.
To deduce $a(N,1)=\alpha_0$, it remains to prove that for any $\alpha\ne \alpha_0$
(with denominator $N$) there exists $j_0$ such that
\begin{equation}\label{qfib}
q_{j_0}
\le 
\big(\alpha_{j_0+1}'q_{j_0}+q_{j_0-1}\big)\alpha_0.
\end{equation}
If $\alpha<\alpha_0$, it holds trivially for $j_0=0$ because 
$\alpha_1'=1/\alpha>1/\alpha_0$. We consequently assume $\alpha>\alpha_0$
and then $a_1=2$. 
If $\alpha$ has a partial quotient which is at least~$3$ then Eq.~\eqref{qfib} holds true because $\alpha_0>1/3$ and we can take $\alpha_{j_0+1}'=[3;\dots]\ge 3$.
Otherwise, the partial quotients of the $\alpha_j'$ are~$1$ or~$2$.
Take $j_0$ such that $a_j=1$ for $1<j\le j_0$, then  by the recurrence formulas Eq.~\eqref{mform}
$q_{j_0}=F_{j_0+2}$
and
$q_{j_0-1}=F_{j_0+1}$. Clearly $\alpha_{j_0+1}'$ is
of the form 
$[2;1,\dots]$,
$[2;2,\dots]$
or
$[2;2]=[2;1,1]$. In any of these cases $\alpha_{j_0+1}'>7/3$ and the right hand side of~Eq.~\eqref{qfib} is greater than
\begin{equation}
\alpha_0\big(
\frac{7}{3}F_{j_0+2}+F_{j_0+1}
\big)
=
\alpha_0
q_{j_0}
\big(
\frac{7}{3}+\frac{F_{j_0+1}}{F_{j_0+2}}
\big)
\ge 
\alpha_0
q_{j_0}
\big(
\frac{7}{3}+\frac{F_{2}}{F_{3}}
\big)
=
\frac{17}{6}
\alpha_0
q_{j_0}
\end{equation}
and \eqref{qfib} follows.

\

Revising the proof, one notes that the assumption $N=F_J$ was only employed to compute the value at $\alpha=\alpha_0$. The proof still applies for $N> F_J$ except that $\alpha=\alpha_0$ is not a valid value in~Eq.~\eqref{a_NF}. Then we have proved
\begin{equation}\label{th1}
a(N,1)
\le \frac{F_{J-2}}{F_{J}}
\quad\text{ if}\quad
N\ge F_J,
\quad
\text{with equality if $N=F_J$.}
\end{equation}
The well-known asymptotic
$F_j\sim r^j$
with $r$ the golden ratio,
gives the asymptotic bound $a(N,1)\le r^{-2}=0.381966\dots$, $N\to\infty$. This value is well within the safe region Eq.~\eqref{a_case1} in which there are no tachyonic instabilities. The values of $N$ reaching the bound Eq.~\eqref{th1} appear in the numerical data as outliers because of the exponential growth. As a final comment we translate our result to the original Physics notation by saying that for $N=F_J$ the optimal value corresponds to $k=\bar{k}= F_{J-2}$ and electric flux either $e=1$ or $e=k$ (any of them). If we restrict ourselves to $N$ being a prime number, one should look only at Fibonacci numbers having this property. Indeed, taking  $J$  a small prime number, $N=F_J$ is also prime (it holds for $J<19$, $N<4181$), which could be of practical use, but we do not know of a general rule which would provide a solution to the fourth point in our list.
In fact the existence of infinitely many Fibonacci primes, as other exponential problems (Mersenne primes or Fermat primes), is considered out of reach with current knowledge in number theory.

We turn now to the most important property of our list (point number 2). The problem of finding  general lower bounds is connected  with  {Zaremba's conjecture}. This is a problem posed in~1971 (see an overview in \cite{kontorovich}) that remains open yet. 
It claims the existence of a positive integer $A$ with the following property:
\begin{equation}\label{zaremba}
\text{For every }  N\in\mathbb{Z}^+,
\text{there exist $a_1,a_2,\dots,a_j\le A$ such that }
[0; a_1,a_2,\dots, a_j]= 
\frac{p_j}{N}.
\end{equation}
In a more elementary way, it means that the recurrence $q_{j+1}=a_{j+1}q_j+q_{j-1}$,
$q_0=0$, $q_1=1$ can capture any positive integer with a judicious choice of the $a_j$.

Indeed the lower bound for $a(N,1)$ is a straightforward consequence of the formula Eq.~\eqref{a_err} under Eq.~\eqref{zaremba}, because
\begin{equation}
a(N,1)>
\max_\alpha
\min_{j}
\frac{1}{\alpha_{j+1}'+1}
\ge\frac{1}{A+2}.
\end{equation}
A kind of converse is also true: using
$\alpha_{j+1}'\ge a_j$, if $a(N,1)>\epsilon$ then we could take 
$A=\lfloor \epsilon^{-1}\rfloor$
in Eq.~\eqref{zaremba} for that $N$, where $\lfloor x\rfloor$ means the integral part. 

Recently there was a breakthrough on Eq.~\eqref{zaremba}. 
In \cite{BoKo} it has been proved that $A=50$ is valid for any $N$ except for a zero density set (i.e. if there are $E_N$ exceptions less than $N$, then $E_N/N\to 0$ as $N\to\infty$). 
Unfortunately the resulting bound $1/52$ is small enough not to 
be conclusive about  the absence of tachyonic instabilities. 
Nonetheless, it is thought that $A=2$ for a certain (large value of) $N$ onwards. If this is true, one argument that we do not reproduce here  (essentially bounding $q_j/q_{j-1}>5/4$) would lead to $a(N,1)>5/19\sim 0.26315\ldots$ for large $N$. That would clearly suffice to exclude the necessity of tachyonic instabilities in the large $N$ limit. We must add that we have also explored the question numerically for $N$ one of the first 4000 prime numbers ($N<37831$) and found $a(N,1)> 0.22779$.
Hence, there are reasons to be optimistic.

\section{Conclusions and Outlook}
In this paper we have applied results and methods arising from the
mathematical area of Number Theory to a problem that appears when
studying the behaviour of SU($N$) Yang-Mills fields in 2+1 dimensions where
the spatial dimensions are compactified in a 2-torus with twisted
boundary conditions. The main point is to determine whether it is
possible to choose the integer flux $k$, characterizing these boundary
conditions, in such a way as to avoid a large $N$ phase transition
appearing at specific torus sizes. In the absense of this transition
one can continuously connect the small size region, where perturbation
theory applies, to the large size region, where confinement takes place. 
The question has been analyzed numerically in Ref.~\cite{GPGAO} for
various values of $N$ suggesting that the problem can indeed be
avoided. However, building on the understanding of the system provided
by the numerical work, we faced here the goal of trying to determine
whether the result will continue to hold at arbitrarily large values
of $N$. 

The possibility of a phase transition emerges when studying the
self-energy contribution to the energy levels of the system in
non-vanishing electric flux sectors. This contribution is negative and
indeed it can be seen that for several values of the chromo-magnetic
flux $k$ the system will develop a phase transition with 
condensation of certain modes, a phenomenon named as tachyonic
instability. Next to leading order of perturbation theory predicts
this to happen for all values of $k$, but the conclusion is only
trustworthy whenever the problem occurs at sufficiently small values of the
coupling constant. Our first result, presented in the previous
paragraph, has been to show  that for any prime value of $N$, it is always possible to choose $k$ in such a way as to make the
perturbative prediction of a phase transition inconclusive. This is
based on the lower bound obtained for the quantity $a(N,2)$ defined in the text. 

We then proceeded to study the problem in general. This requires a
certain understanding of the behaviour of the system in the
non-perturbative region. For that purpose we built on the results
obtained in previous numerical studies in which the behaviour of the
energies at all torus sizes is well described by a function involving 
all the parameters of the problem. This function allows us to
extrapolate our analysis to arbitrary values of $N$ and $k$. On the
basis of this, we studied the necessary condition to be able to avoid 
any instability occuring at any value of the torus size and $N$. The
condition takes the form of a lower bound on $a(N,1)$. The existence
of such a lower bound and its actual value turns out to be related to
a conjecture in Number Theory, formulated by Zaremba in
1971~\cite{zaremba} (see also \cite{kontorovich}), and which remains to be proven. The situation is aggravated by the necessity that the bound is large enough to guarantee the avoidance of instabilities. However, on the positive side we must mention that, from the Physics perspective, it would be enough if the bound holds only for large enough $N$, or just for a sequence of values of $N$ running up to infinity. Indeed, one of the results of this paper was to show that there exist an optimal sequence given by the Fibonacci numbers for which the instability can be avoided.  Nonetheless, if we insist that $N$ should be a prime number, to avoid other potential problems, we are faced with the question of existence of an infinite number of primes in the Fibonacci sequence.

To summarize, we can say that although the issue has not been 
definitely settled, our work has established a connection with  interesting open problems in Number Theory, which might eventually lead to a full understanding of this and other related physical results.

\section*{Acknowledgments}
The first author is partially supported by the grant MTM2014-56350-P from the Ministerio
de Econom\'{\i}a y Competitividad (Spain).  A.G-A acknowledges financial support from the grants FPA2012-31686, 
FPA2012-31880, FPA2015-68541-P and the MINECO Centro de Excelencia Severo Ochoa Program SEV-2012-0249. 
We thank Margarita Garc\'{\i}a P\'{e}rez for discussions and a critical reading of the manuscript. We also thank the authors of Ref.~\cite{prepar} for allowing us to use some of their data prior to publication.

\bibliography{paper.bbl}{}
\bibliographystyle{JHEP}

\end{document}